\documentclass[prl,twocolumn,superscriptaddress,showpacs]{revtex4}

\usepackage{amsbsy}
\usepackage{amsfonts}
\usepackage{amssymb}
\usepackage{amsmath}
\usepackage{color}
\usepackage{graphicx}
\usepackage{float}
\usepackage[caption = false]{subfig}

\begin{document}

%opening

\title{A quantum rectifier in a one-dimensional photonic channel}

\author{E. Mascarenhas}
\affiliation{Laboratory of Theoretical Physics of Nanosystems, Ecole Polytechnique F\'{e}d\'{e}rale de Lausanne (EPFL), CH-1015 Lausanne, Switzerland}

\author{M. F. Santos}
\affiliation{Departamento de F\'{i}sica, Universidade Federal de Minas Gerais, CP 702, 30123-970 Belo Horizonte, Brazil}

\author{A. Auff\`{e}ves}
\affiliation{CNRS and Universit\`{e} Grenoble Alpes, Institut N\'{e}el, F-38042, Grenoble, France}

\author{D. Gerace}
\affiliation{Dipartimento di Fisica, Universit\`{a} di Pavia, via Bassi 6, I-27100 Pavia, Italy}

\begin{abstract}
By using a fully quantum approach based on an input-output formulation of the stochastic Schr\"{o}dinger equation, we show rectification of radiation fields in a one-dimensional waveguide doped with a pair of ideal two-level systems for three topical cases: 
classical driving, under the action of noise, and single photon pulsed excitation. 
%We show that classical approximations may overestimate the rectification efficiency. 
We show that even under the constant action of unwanted noise the device still operates effectively as an optical isolator, which is of critical importance for noise resistance. Finally, harnessing stimulated emission allows for non-reciprocal behavior for single photon inputs, thus showing purely quantum rectification at the single-photon level.
The latter is a considerable step towards the ultimate goal of devising an unconditional quantum rectifier for arbitrary quantum states.
\end{abstract}

%\pacs{ 03.65.Yz; 03.67.Hk; 03.67.Pp}

\maketitle

\textit{Introduction}. One of the most viable perspectives for future information and communication technologies relies on exploiting photonic transport in integrated circuits, which have been experiencing progressive technological achievements and miniaturization \cite{focus}.
In this context, there is a growing need for the design and implementation of novel optical devices that are able to control photonic transport. Photonic rectification, or optical isolation, lies at the core of such ongoing efforts, with the main aim of devising and ultimately realizing systems that are able to make light propagation unconditionally unidirectional \cite{isolation}. The impact over current research in nanophotonics is potentially at the level of the role played by electronic diodes. Furthermore, at low temperatures the conduction of heat is mainly executed by photons~\cite{ThermoF} and preventing unwanted heating is of paramount importance in quantum processing. 
In fact, such \textit{photonic diodes} represent key elements to prevent undesired fields to propagate backwards to strategic centers in prospective photonic integrated circuits, thus preserving the processing capabilities of the sources of signals. Such circumstances are required to avoid decoherence within the circuit, by ensuring the signals to be efficiently transmitted to their targets, and information processing units not to be disturbed by undesired reflections.

After an initial focus on rectification of classical driving signals \cite{gallo2001,Lepri,yu2009nphot,Bi2011nphot,Lifan2012,Lira2012,alu2014}, these concepts are now transferring to the realm of quantum information, where nonlinear systems at the single quantum level are likely to play a major role. Previous proposals have already addressed the possibility of achieving a unidirectional behavior from low energy coherent input radiation fields \cite{Diode,FP}. Quantum rectification, i.e. non-reciprocal transmission of quantum states of light, has been mainly addressed for optically active media in circularly polarized waveguides \cite{shen2011prl}. However, an unconditional and passive rectifier of single-photon wave packets has not been designed and rectification in photonic circuits is still in its infancy. Here we take considerable steps in the design of an efficient, noise resistant rectifier operating both in the classical and in the purely quantum regime.
We model the one dimensional doped waveguide with a fully quantum input-output stochastic theory. 
First, we revisit a recently proposed one-dimensional configuration, where a Fabry-Perot cavity made of two ideal two-level systems simultaneously provides rectification and high transmission for coherent input states \cite{FP}. 
The system is shown to be resilient to noise still providing mensurable optical isolation even under extenuating circumstances in which the system is simultaneously driven with the desired signal and with an unwanted noisy field. Most importantly, we present yet another novel finding: harnessing stimulated emission allows for non-reciprocal behavior for single photon inputs, thus showing purely quantum rectification at the single-photon level. Our results are of paramount importance for upcoming quantum technologies in which it is highly desirable to control light transport at the single photon level, that is, at the level of the fundamental transporters of information and heat.

\begin{figure}
{\includegraphics[width = 8cm]{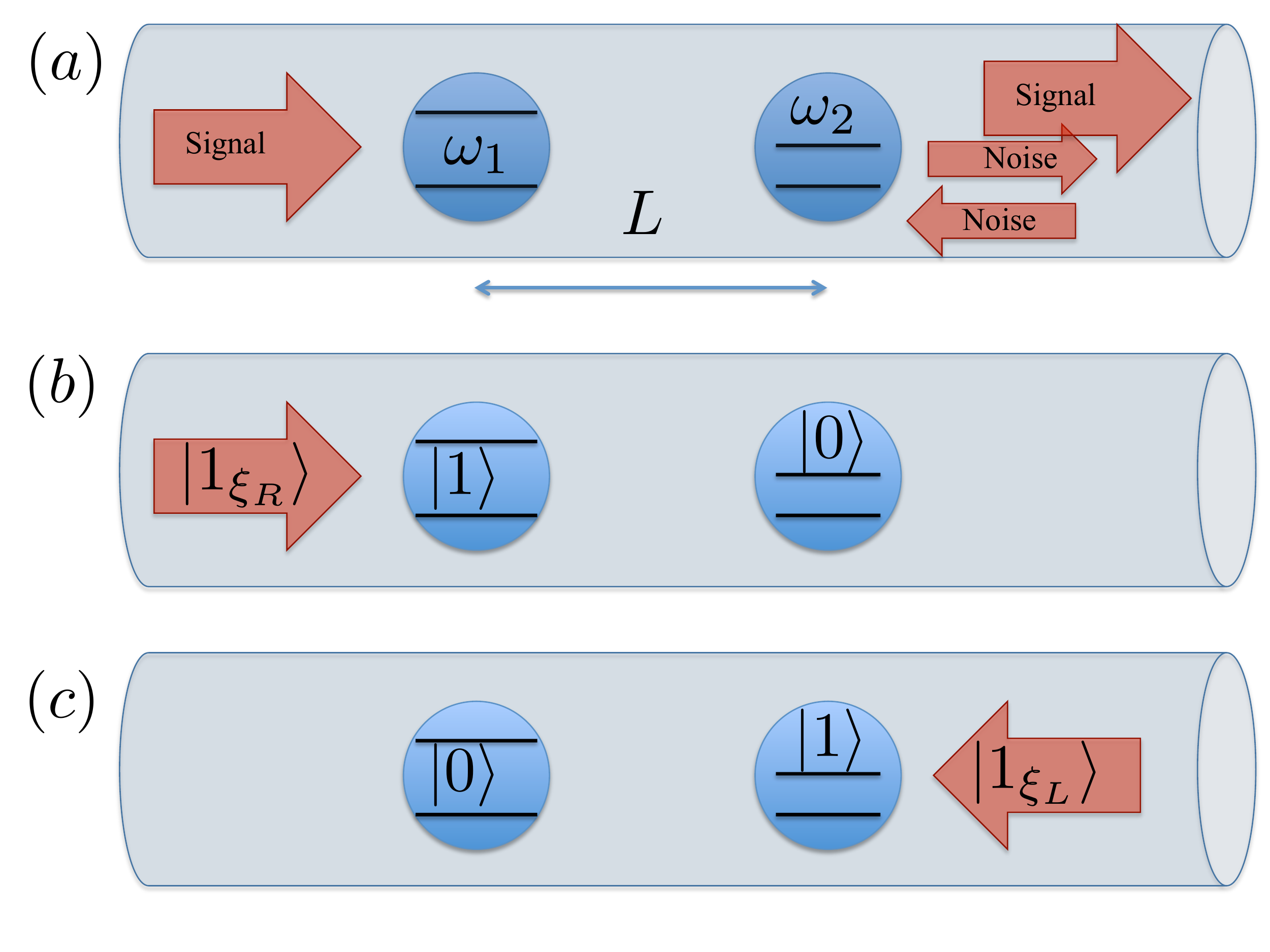}}\\ 
\caption{Schematic description of a one-dimensional channel with a pair of two-level quantum systems spatially separated by a distance $L$, with ground-to-excited state transition resonances $\omega_1$ and $\omega_2$. $(a)$ The device working resiliently by simultaneously transmitting a signal while blocking noise. (b) and (c) correspond to the initial conditions of the single photon input $|1_{\xi_{R(L)}}\rangle$ coming from the left (right) with time profile $\xi_{R(L)}(t)$ when the atom first receiving the input is inverted to its excited state.}
\label{Scheme}
\end{figure}

\textit{Theory}. 
The one-dimensional photonic device under investigation is schematically shown in Figure~\ref{Scheme}: it is a closed channel for radiation propagation coupled to two-level systems. Such an ideal system can be realized in different solid state integrated platforms, involving either semiconductor \cite{Lodhal,Claudon}, metallic \cite{Akimov}, or superconductor \cite{VanLoo} materials technology, where artificial atoms can be treated as externally tunable and close-to ideal two-level systems. 
Due to the presence of the two quantum emitters, a fully quantum approach is essential to correctly capture the dynamical evolution and stationary state of this system.
We model the one-dimensional device through a generalized quantum master equation formulation of input-output theory, which was recently extended to arbitrary one-dimensional closed channels made of a sequence (array) of elementary quantum systems~\cite{Chang1,Chang2}. First, we reformulate this powerful approach in the framework of a quantum stochastic 
Schr\"{o}dinger equation, which allows us to obtain a more straightforward determination of the output currents. 
The two infinite one-dimensional channels are explicitly described as left and right modes, respectively, to which we attribute annihilation operators $\hat{r}_{\omega}$ and $\hat{l}_{\omega}$. Hence, the full Hamiltonian describing the interaction between the waveguide modes and the quantum elements of the array is given by
\begin{equation}H=\sum_i\int d\omega \sqrt{\gamma_{\omega}}\hat{\sigma}_i^{\dagger}\left[ \hat{r}_{\omega}e^{i\omega(x_i/c-t)}+ \hat{l}_{\omega}e^{-i\omega(x_i/c+t)}\right]+\mathrm{h.c},\end{equation}
with $\sigma_i$ being an annihilation operator for the $i$th site of the array, and $x_i$ is the corresponding spatial coordinate. By making the standard Markovian approximations assuming flat spectral coupling, the corresponding Heisenberg-Langevin and density matrix master equations were already derived and extensively presented in Refs.~\onlinecite{Chang1,Chang2}. 

Here we recast this model into quantum stochastic calculus for input-output fields~\cite{Qnoise}, similarly to the single qubit case already presented in Ref.~\onlinecite{Fan}. The main advantage of this approach is that the Markovian approximation is straightforwardly encoded in the formalism through the Ito product rules~\cite{Qnoise}.
Thus, we define the jump operators
\begin{equation}
\hat{J}_{R}=\sqrt{\gamma}\sum_i e^{-i\omega_{\mathrm{in}}x_i/c} \hat{\sigma}_i, \quad \mathrm{and}\quad 
\hat{J}_{L}=\sqrt{\gamma}\sum_ie^{i\omega_{\mathrm{in}}x_i/c} \hat{\sigma}_i  \, ,
\end{equation}
which can be used to recast the problem into a stochastic Schr\"{o}dinger equation for the state of the waveguide and two-level systems in the Stratonovich form~\cite{Qnoise}
\begin{multline}d|\Psi\rangle=\Bigg[\hat{J}_R \hat{dR}^{\dagger}+ \hat{J}_L \hat{dL}^{\dagger} -
\hat{J}_R^{\dagger} \hat{dR} - \hat{J}_L^{\dagger} \hat{dL} \\
-i\gamma\sum_i \sin{(k |x_i-x_j|)}(\hat{\sigma}_i^{\dagger} \hat{\sigma}_j dt +\mathrm{h.c.})\Bigg]|\Psi\rangle \, , 
\end{multline}
with $k=\omega_{\mathrm{in}}/c$ describing the input wave vector, and the quantum Wiener noise term being formally expressed as $\hat{R}(t)=\int_0^tr(s)ds$ with $\hat{r}(t)=\int \hat{r}_{\mathrm{\omega}}e^{-i\omega t}d\omega$ (and analogously for $\hat{L}$), where vacuum Ito rules are $\hat{dR}\hat{dR}^{\dagger}=\hat{dL} \hat{dL}^{\dagger}=\openone dt$. 
This leads to the Ito equation 
\begin{multline}
d|\Psi\rangle=\Bigg[\hat{J}_R \hat{dR}^{\dagger}+\hat{J}_L \hat{dL}^{\dagger}-\hat{J}_R^{\dagger} \hat{dR}- \hat{J}_L^{\dagger} \hat{dL} \\
-i\gamma\sum_i \sin{(k |x_i-x_j|)}(\hat{\sigma}_i^{\dagger}\hat{\sigma}_jdt +\mathrm{h.c.})\Bigg]|\Psi\rangle \\
-\frac{1}{2}(\hat{J}_R^{\dagger} \hat{J}_R+ \hat{J}_L^{\dagger} \hat{J}_L)|\Psi\rangle dt \, .
\end{multline}
By making use of the stochastic chain differentiation, $d\rho=d|\Psi\rangle\langle\Psi|+|\Psi\rangle d\langle\Psi|+d|\Psi\rangle d\langle\Psi|$, we can finally derive the stochastic master equation
\begin{multline}
d\rho=-i\left[ H_0,\rho \right]dt +\Bigg[\hat{J}_R \hat{dR}^{\dagger}+\hat{J}_L \hat{dL}^{\dagger}-\hat{J}_R^{\dagger} \hat{dR}- \hat{J}_L^{\dagger} \hat{dL} \\
-i\gamma\sum_i \sin{(k |x_i-x_j|)}(\hat{\sigma}_i^{\dagger} \hat{\sigma}_j dt +\mathrm{h.c.}), \ \rho\Bigg] \\
-\frac{1}{2}\left[\mathcal{D}_R(\rho)+\mathcal{D}_L(\rho)\right]dt \, ,
\end{multline}
where the dissipative operators are defined as 
$\mathcal{D}_{R,L}(\rho)=\hat{J}_{R,L}^{\dagger} \hat{J}_{R,L}^{\dagger}\rho+\rho \hat{J}_{R,L}^{\dagger} \hat{J}_{R,L}^{\dagger} -2\hat{J}_{R,L}\rho \hat{J}_{R,L}^{\dagger}$, 
and $H_0$ is the system Hamiltonian describing its internal dynamics.

We will consider two types of input states. A purely classical state, such as a continuous wave (cw) coherent pump, and a purely quantum one, i.e. single photon pulse, respectively. 
For the case of cw coherent pump we have $\langle \hat{dR} \rangle=\mathcal{E} dt$, where the light-matter detuning is included in $H_0=\sum_i\Delta_i \hat{\sigma}^{\dagger}_i \hat{\sigma}_i$, and $\Delta=\omega_i-\omega_{\mathrm{in}}$ in a rotated reference frame. Whereas, for the single photon input we use the formulation discussed in Refs.~\onlinecite{Single,Andre}. 
%It is important to stress that the approach proposed here should only be valid in the case that the photon spatial extension is larger than the separation between the quantum systems, i.e. the photon wave packet is spread over the entire array when being transmitted. In the present case this is particularly convenient, since our interest is focussed on an array of only two subsystems. 
%Otherwise, if the packet is too short the interaction with the subsystems can be regarded as independent events and this regime leads to no rectification. 
%We have performed classical simulations for two Gross-Pitaevskii-like nonlinear defects and have observed that when the defects are far apart the system
%is not a good rectifier for short pulses in the spirit of~\cite{Lepri} (not show here).
In this case, the master equation for the density matrix can be projected onto the single photon occupation Fock state, $|1_{\xi_R}\rangle$, with $\hat{dR}|1_{\xi_R}\rangle=\xi_Rdt|0\rangle$, where the  normalization condition can be imposed as $\int_0^{\infty}|\xi_R(t)|^2dt=1$. This leads to the following set of coupled master equations
\begin{equation}
\dot{\rho}_{11}=\mathcal{L}\rho_{11}+[\rho_{01},\hat{J}_{R,L}^{\dagger}]\xi_{R,L}+[\hat{J}_{R,L},\rho_{10}]\xi_{R,L}^{\ast},
\end{equation}
\begin{equation}\dot{\rho}_{01}=\mathcal{L}\rho_{01}+[\hat{J}_{R,L},\rho_{00}]\xi_{R,L}^{\ast},
\end{equation}
and $\dot{\rho}_{00}=\mathcal{L}\rho_{00}$, $\mathcal{L}$ being the Lindblad operator that accounts for the system internal Hamiltonian and both coherent and incoherent couplings with the one-dimensional channels. 
This set of coupled equations is to be solved with initial conditions $\rho_{ij}=\rho_{\mathrm{initial}}\delta_{ij}$ noting that each $\rho_{ij}=\langle i_{\xi_{R,L}}|\rho|j_{\xi_{R,L}}\rangle$ is a 4x4 matrix.

\begin{figure}
{\includegraphics[width = 3.2in]{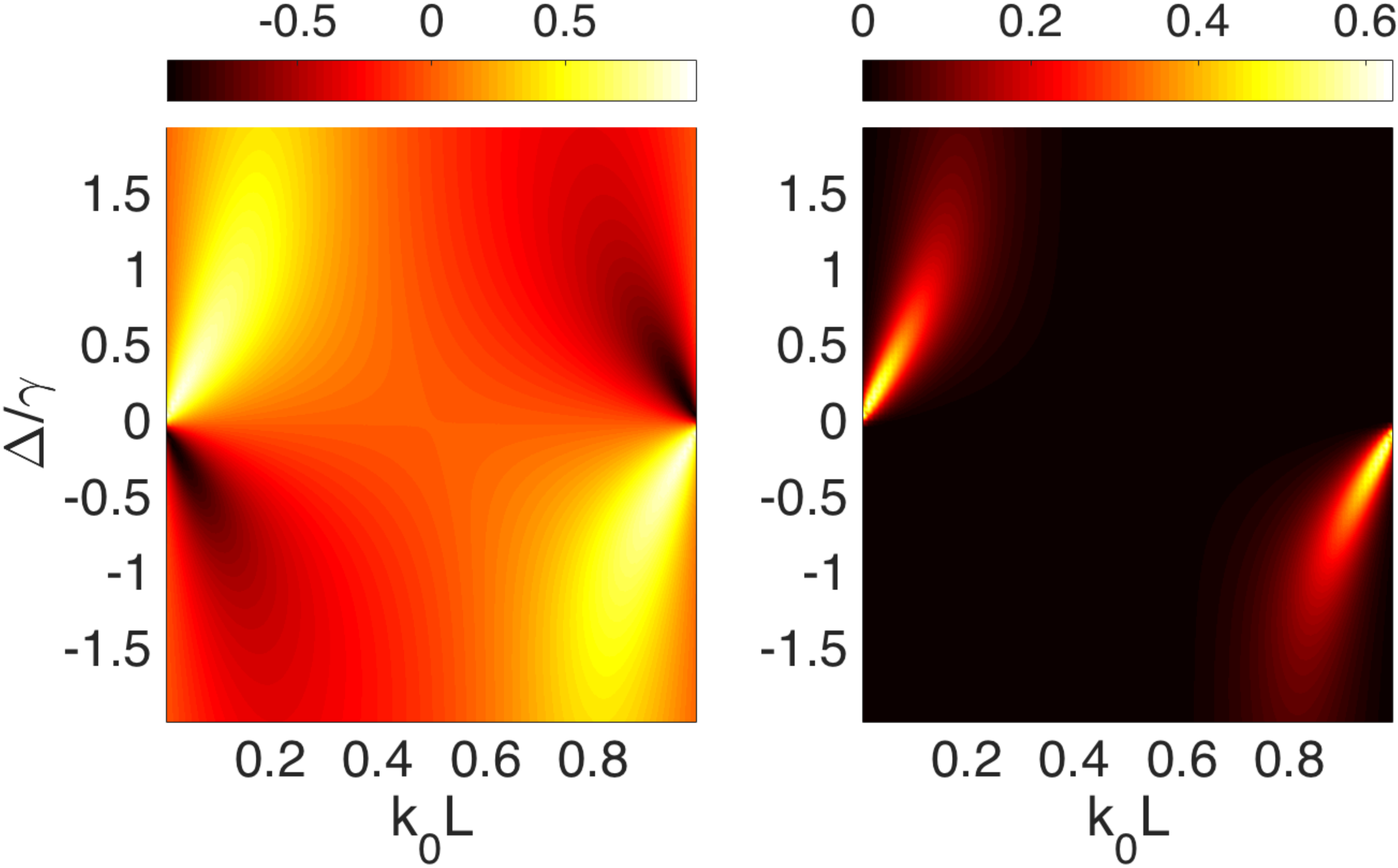}}\\ 
\caption{(Left) The rectification factor under unidirectional pumping and (Right) the corresponding diode efficiency as functions of detuning $\Delta_1=\Delta$ ($\Delta_2=0$) and interatomic distance $L$. The pump strength was partially optimised to reach higher diode efficiency and it was set to $\mathcal{E}=\sqrt{0.05}\gamma$.}
\label{Rect}
\end{figure}

Within this framework, the output fields are defined as $\hat{dR}_{\mathrm{out}}=\hat{dR}+\hat{J}_R dt$ and $\hat{dL}_{\mathrm{out}}=\hat{dL}+\hat{J}_Ldt$, which corresponds to evaluating the fields outside the system of interest in a position that eliminates the complex spatial phase of the input field, which can always be done. 
The output photon occupation numbers can also be derived by using standard input-output theory~\cite{Qnoise}. 
Hence, the equation of motion for the output photon process is given by
\begin{equation} 
\hat{dN}_{R\mathrm{out}}=\hat{dN}_R+\hat{J}_R \hat{dR}^{\dagger}+\hat{J}_R^{\dagger} \hat{dR}+ \hat{J}_R^{\dagger} \hat{J}_R dt \, .   
\end{equation}
In the special case of a single photon input we get the average value 
\begin{multline} 
\langle \hat{dN}_{R\mathrm{out}}\rangle_{11}=|\xi_R|^2 dt+(\langle \hat{J}_R \rangle_{10}\xi_R^{\ast}+\langle \hat{J}_R^{\dagger}\rangle_{10} \xi_R) dt \\
+ \langle \hat{J}_R^{\dagger} \hat{J}_R \rangle dt \, .   
\end{multline}

\textit{Results}. We start by evaluating the rectification properties of the system under the continuous wave pump.
We define the rectification factor under unidirectional pump along the same lines as it was previously done in Refs.~\onlinecite{Lepri,Diode,FP}, explicitly as a function of the output currents
\begin{equation}
\mathcal{R}=
\frac{\langle \hat{N}_{R\mathrm{out}}(k)\rangle-\langle \hat{N}_{L\mathrm{out}}(-k)\rangle}
{\langle \hat{N}_{R\mathrm{out}}(k)\rangle+\langle \hat{N}_{L\mathrm{out}}(-k)\rangle}.
\end{equation}
The rectification factor measures the unbalance between the two directions of transmission when the system is pumped \emph{either} from the left \emph{or} from right. 
The total photonic diode efficiency is thus the product between the rectification factor and the transmission coefficient, $D=\mathcal{R} T_R$, where the transmission coefficient is defined as $T_R=\langle \hat{N}_{R\mathrm{out}}(k)\rangle/\langle \hat{N}_{R}(k)+\hat{N}_{L}(k)\rangle$.

In Fig.~(\ref{Rect}) we show both the rectification factor and the photonic diode efficiency. 
We set the pair of two-level systems in the same configuration as in Ref.~\onlinecite{FP}, i.e. the right atom is kept in resonance with the input light field while the left atom may be detuned.
By comparing such results with the semiclassical analysis performed in Ref.~\onlinecite{FP} we see that the Fabry-Perot approach overestimates the diode efficiency, which is always below $70\%$ when a fully quantum analysis is performed.

\begin{figure}
{\includegraphics[width = 3.3in]{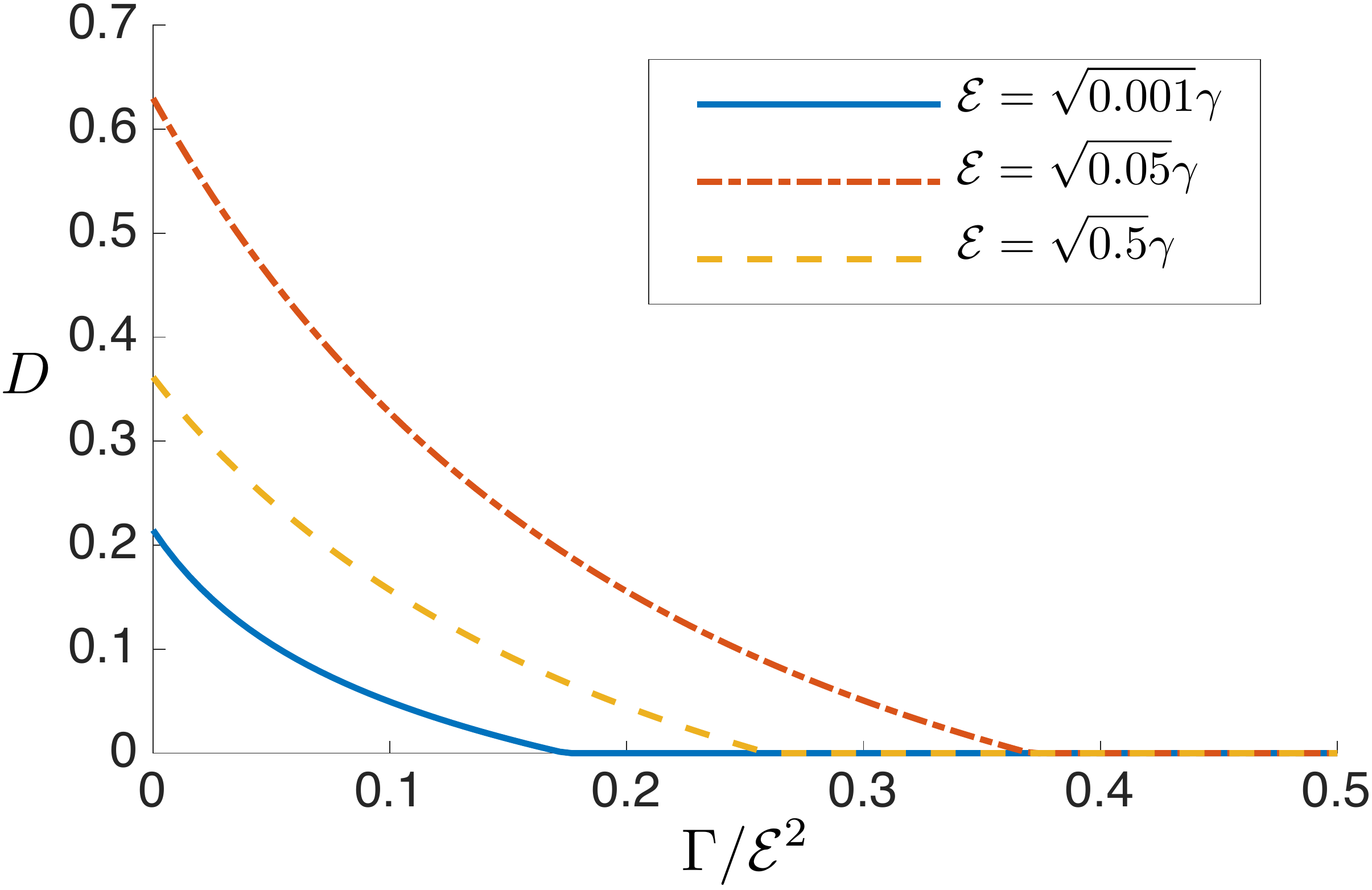}}\\ 
\caption{The diode efficiency optimized over $\Delta$ and $L$ under the action of noise as sketched in Fig.~\ref{Scheme}-a.}
\label{TwoSidePump03one10}
\end{figure}

\textit{Rectification under noise}. In the perspective of estimating the noise resistance of the device, it is important to investigate if it still rectifies when it is \emph{simultaneously} pumped from both sides (see Fig. \ref{Scheme}-a). This is typically the case in realistic situations where a signal is injected into the device, while noise is blocked by the device. 
If the unwanted field is weak compared to the signal then the net rectification should not be significantly altered. We assume a fluctuating input traveling in the left going mode. More specifically, we assume $\langle dL\rangle=dW$ with $W$ being a classical Wiener noise with moments $\overline{dW}=0$ and $\overline{dW^2}=\Gamma dt$ such that we have a noisy input in the left going mode with the intensity $\Gamma$. This corresponds to a fluctuating field with a broad and unstructured spectrum. In order to obtain the average state we expand the Hamiltonian evolution up to second order using Ito calculus and then average over the classical noise. This leads to the following extra term in the master equation 
\begin{equation}d\rho=-\frac{\Gamma}{2}\left[ X_L^2\rho+\rho X_L^2 -2X_L\rho X_L\right],\end{equation} 
which can be interpreted as a global dephasing process with $X_L=i(J_L-J_L^{\dagger})$ and adds an input to the system $\langle dN_L\rangle=\Gamma dt$. In Fig.~(\ref{TwoSidePump03one10}) we show the diode efficiency optimized over the signal frequency and inter-atomic distance as a function of the noise-signal ratio $\Gamma/\mathcal{E}^2$.

\begin{figure}
{\includegraphics[width = 3.3in]{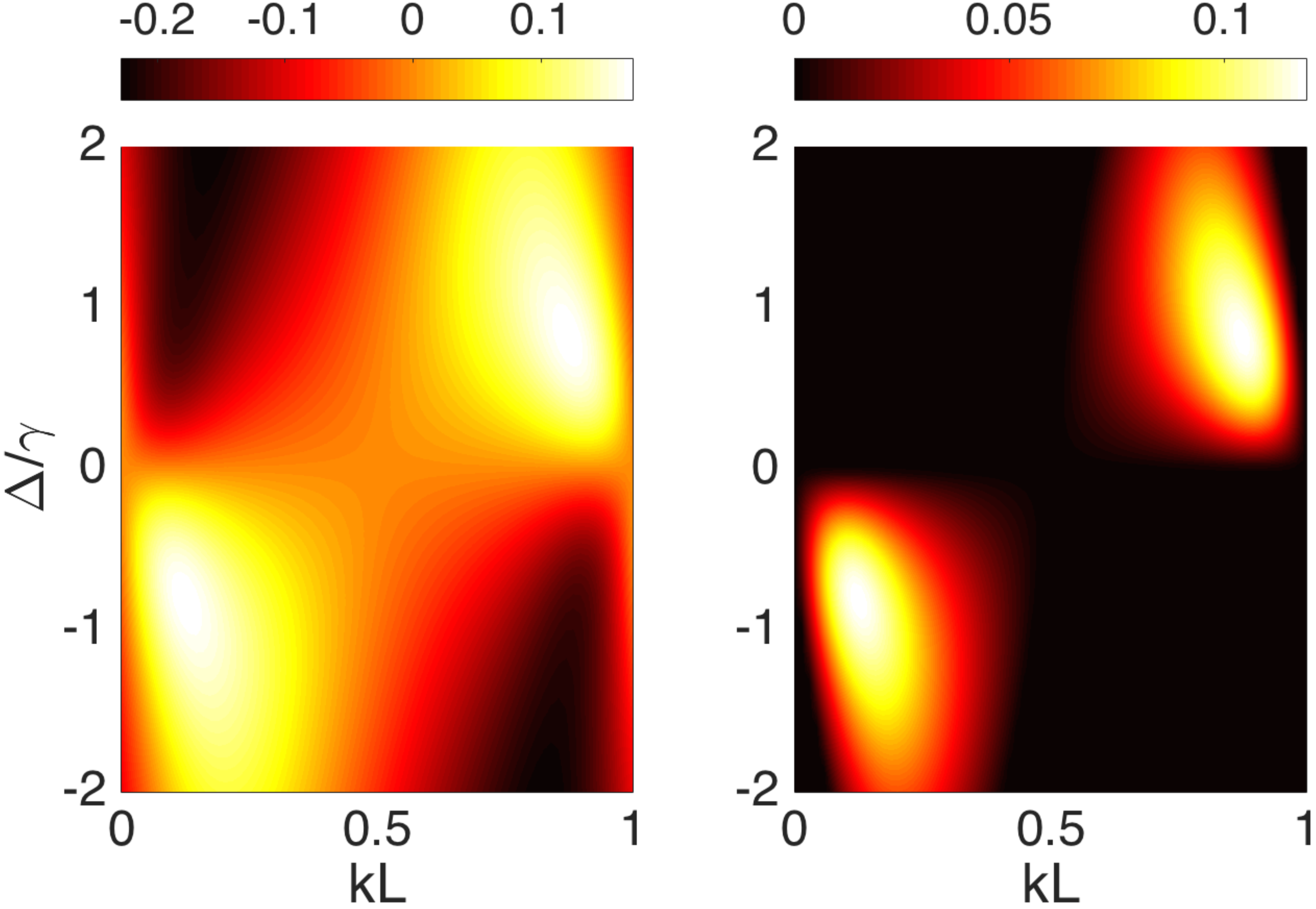}}\\ 
\caption{The rectification $\mathcal{R}$ and the diode efficiency $\mathrm{max}(D,0)$ with $D=\mathcal{R}T_R$ and $T_R=N_{R\mathrm{out}}/N_R$ under unidirectional single photon input $|1_{\xi_R(L)}\rangle$, with $\xi_{R(L)}\propto e^{-\Gamma t/2}$ and $\Gamma=2\gamma$ when the two-level system receiving the input is in the excited state. In this case we take $N_R=2$ rather than $N_R=1$ in order to take into account the extra excitation in the inverted atom. This guarantees that the net diode efficiency is not overestimated.}
\label{SingleP}
\end{figure}

\textit{Single Photon Rectification through Stimulated Emission}. As a final case, we now discuss the quantum rectification of a single-photon input pulse by the stimulated emission of an excited two-level system. 
Firstly, we point out that, being a non-linear effect, one should not expect rectification in the single excitation manifold where the system effectively responds linearly to the pump. This is confirmed by the fully symmetric transmission coefficient with respect to exchanging the atomic frequencies, as obtained in~\cite{Baranger} [see equation (S$4$) in the supplementary material]. 
To rectify a single photon input we exploit optimal irreversible stimulated emission~\cite{Dani}. We initially prepare the emitter receiving the input in its excited state while the other is left in its ground state as shown in figure~(\ref{Scheme}-$b$) and~(\ref{Scheme}-$c$). The single photon input tends to stimulate the emission
of the excited atom into the same mode and hence in the same direction of propagation. Combining stimulated emission with the asymmetric frequency disposition of the emitters yields rectification under the single photon input. 
We assume exponentially decaying wave packets, $\xi(t)\propto e^{-\Gamma t/2}$, as they are naturally generated by the emission of a single photon from another two-level system and such packets are optimal in stimulating the emission~\cite{Dani}. The results for the rectification under such conditions are shown in Fig.~(\ref{SingleP}).As it appears from the plot, sensible rectification of single propagating photons is obtained with our setting of the device, owing to the non-linearity of stimulated emission. In Fig.~(\ref{SingleP}-b) we focus on the efficiency of the diode, where the transmission is now computed with $N_R=2$ rather then $N_R=1$, to take into account the extra-excitation used to prepare one of the atoms in its excited state. It shown that the diode is able to block the optimal stimulated emission~\cite{Dani} to the left while it still transmits to the right.

\textit{Summary}. We have presented a fully quantum model to investigate the performances of a noise resistant, quantum rectifier. Under classical pumping, our device is shown to provide optical isolation even in the presence of unwanted noise of the same order of magnitude as the signal to be transmitted. On the other hand, exploiting stimulated emission allows reaching substantial rectification for single propagating photons. Such configuration is promising for high fidelity quantum rectification. As future steps we envision redundant input states such that several copies of the desired signal state are provided as a single input. Such input states would be rectified and an error correction scheme may be designed such that out of the several imperfect transmitted copies a single high fidelity copy my be distilled~\cite{Distil}. Therefore, the present contribution is likely to provide an important step towards unconditional high fidelity quantum state rectification.

E.M acknowledges Daniel Valente, Vincenzo Savona and Filippo Fratini for inspiring conversations on the subject.
D.G. acknowledges support from the Italian Ministry of University and Research (MIUR) through FIRB Futuro in Ricerca Project No. RBFR12RPD1, and the Brazilian CNPq through the Special Visiting Researcher program PVE/CSF, Project No. 407167/2013-7.

\end{document}